\documentclass{elsart}

\usepackage{graphics}

\usepackage{amssymb}
\usepackage{amsmath}

\newenvironment{proof}[1][Proof]{\noindent\textbf{#1.} }{\ \rule{0.5em}{0.5em}}

\newcommand{\qexp}{ \exp_{q} }
\newcommand{\qln}{ \ln_{q } }
\newcommand{\qprod}{ \mathop{\otimes}_{q} }
\newcommand{\kln}{ \ln_{\{ \kappa \}} }
\newcommand{\kexp}{ \exp_{ \{\kappa \}} }
\newcommand{\kprod}{ \mathop{\otimes}_{\kappa} }
\newcommand{\aprod}{ \mathop{\otimes}_a }

\newcommand{\Sk}{ S_{ \kappa } }

\begin{document}
\begin{frontmatter}

\title{$\kappa$-generalization of Gauss' law of error}

\author[wada]{Tatsuaki Wada\corauthref{cor1}} and
\ead{wada@ee.ibaraki.ac.jp}
\author[suyari]{Hiroki Suyari}
\ead{suyari@faculty.chiba-u.jp, suyari@ieee.org}
\address[wada]{Department of Electrical and Electronic Engineering, 
Ibaraki University, Hitachi, Ibaraki 316-8511, Japan}
\address[suyari]{Department of Information and Image Sciences, 
Faculty of Engineering, Chiba University, 263-8522, Japan}
\corauth[cor1]{Corresponding author.}
\begin{abstract}
Based on the $\kappa$-deformed functions ($\kappa$-exponential
and $\kappa$-logarithm) and
associated multiplication operation ($\kappa$-product) introduced
by Kaniadakis (Phys. Rev. E \textbf{66} (2002) 056125), we present 
another one-parameter generalization of Gauss' law of error. 
The likelihood function in Gauss' law of error is generalized
by means of the $\kappa$-product. This $\kappa$-generalized
maximum likelihood principle leads to the {\it so-called} $\kappa$-Gaussian
distributions.
\end{abstract}

\begin{keyword}
Gauss' law of error \sep $\kappa$-deformed function \sep $\kappa$-product 
\PACS 02.50.Cw \sep 05.20.-y \sep 06.20.Dk
%
\end{keyword}
\end{frontmatter}

\section{Introduction}
\vspace*{-7mm}
Gaussian (or normal) distribution \cite{Jaynes} is one of the most 
well-known and fundamental distributions in many fields of science,
e.g., an error distribution in measurement, a probability distribution
of a fluctuating physical quantity in statistical mechanics, which accounts 
for the successes of Boltzmann-Gibbs exponential distributions \cite{Reif}.
For statistically independent $N$ events, it is well-known that the limiting
stable distribution becomes a Gaussian one for sufficiently large $N$. This
is the consequence of {\it central limit theorem} \cite{Feller}, which is 
the most famous
theorem in mathematical probability theory. 
Historically however the Gauss derivation, as known as Gauss' law of error,
played a very important role before establishing the central limit theorem.

Quite recently one of the authors {\it et al.} \cite{Suyari} have 
shown a one-real-parameter ($q$) generalization of Gauss' law of error,
in which the $q$-generalized maximum likelihood principle leads to 
the {\it so-called} $q$-Gaussian distribution, by utilizing
the Tsallis $q$-deformed functions \cite{Tsallis88,NEXT2001,NEXT2003,Santa Fe} 
and associated multiplication
operation, i.e., {\it so called} $q$-product \cite{Borges04,Nivanen03}.
Tsallis' entropy and $q$-deformed functions have been successfully
applied in order to explain the ubiquitous existence of power-law behaviors 
in nature. 
Tsallis' entropy $S_q = (\int dx f(x)^q - 1)/(1-q)$ is a one-real-parameter 
generalization of 
Boltzmann-Gibbs-Shannon entropy $S^{\rm BGS} = -\int dx\; f(x) \ln f(x)$ and 
employed for generalizing the 
traditional Boltzmann-Gibbs statistical mechanics, thermodynamics, 
and Jaynes' information theory.\\
The $q$-deformed exponential and logarithmic functions are defined by
\begin{eqnarray}
\qexp(x) &\equiv& \left[ 1+(1-q) x \right]_{+}^{\frac{1}{1-q}}, \\
\qln(x) &\equiv&  \frac{x^{1-q} - 1}{1-q},
\end{eqnarray}
respectively.
Here $q$ is a real parameter characterizing 
the deformation functions, and $[x]_{+} \equiv \max(x, 0)$.
In the $q \to 1$ limit, the $q$-exponential and $q$-logarithmic functions
reduce to the standard exponential and logarithmic functions, respectively.
The following useful identity holds.
\begin{equation}
\exp_q(x + y)  =  \exp_q(x) \cdot \exp_q \left( \frac{y}{1+(1-q)x} \right).
\label{q-identity}
\end{equation}
As in the general usage for a Gaussian function, we here mean 
that any function of the general form
$\qexp(-\alpha x^2)$ with a positive constant $\alpha$ as a 
$q$-Gaussian function.\\
One of the most fundamental ingredients for proving the $q$-generalization 
of Gauss' law of error is the recently introduced 
$q$-product \cite{Borges04,Nivanen03}, which is defined by
\begin{equation}
    x \qprod y \equiv 
     \Big[ x^{1-q} + y^{1-q} -1 \Big]_{+}^{\frac{1}{1-q}}.
\end{equation}

On the other hand,
Kaniadakis \cite{Kaniadakis01,Kaniadakis02} has introduced another 
one-parameter deformed exponential function
\begin{equation}
\kexp(x)  \equiv \left( 
            \sqrt{1+\kappa^{2}x^{2}}+\kappa x \right)^{\frac{1}{\kappa }},
\end{equation}
and its inverse function
\begin{equation}
      \kln (x)  \equiv \frac{ x^{\kappa}-x^{-\kappa }}{2 \kappa }, 
\end{equation}
where $\kappa$ is a real parameter and takes a value in the
range $(-1, 1)$. Both $\kappa$-deformed functions are symmetric
under interchange of $\kappa \leftrightarrow -\kappa$.
In the $\kappa \to 0$ limit, the $\kappa$-exponential and 
$\kappa$-logarithmic functions reduce to the standard exponential and 
logarithmic functions, respectively.\\
Note that the $\kappa$-logarithmic function can be expressed
in terms of the two $q$-logarithmic functions
with the different $q$ indices as follows.
\begin{equation}
   \kln (x) = \frac{1}{2} \ln_{1+\kappa}(x) + \frac{1}{2} \ln_{1-\kappa}(x). 
\end{equation}
However the $\kappa$-exponential function cannot be expressed in terms
of the $q$-exponential functions.\\
Note also that both $q$- and $\kappa$-exponentials shows 
asymptotic power-law behaviors for large $x$:
\begin{align}
   \qexp(x) & \mathop{\sim}_{x \to \infty} 
           \left( (1-q) x \right)^{\frac{1}{1-q}},\\
   \kexp(x) &\mathop{\sim}_{x \to \infty} 
            \left( 2 \kappa x \right)^{\frac{1}{\kappa }},
\end{align}
whereas for a small value of $x$ both functions behave as the
standard exponential functions.
Only in the intermediate region they are slightly different each other.
Although the difference is not so large, this means that if an experimental
data is well fitted with $\kappa$-exponential function it should not be
well fitted with $q$-exponential one, and vice versa.
Until now in the literature there are only a few experimental 
evidences which are well fitted with $\kappa$-exponential probability 
distributions: the flux distribution of the cosmic rays extends
over 13 decades in energy \cite{Kaniadakis}; the rain events 
in meteorology (the number density of rain events versus 
the event size) \cite{Kaniadakis,rain}; and the analysis of the fracture
problem (the relation between the length of a transversal cut of 
the conducting thin ribbon and the electrical resistance) \cite{Cravero04}.

Kaniadakis \cite{Kaniadakis} had already introduced a deformed 
algebra based on his $\kappa$-deformed functions before establishing
the concept of the $q$-product \cite{Borges04,Nivanen03}.
His $\kappa$-deformed product \cite{Kaniadakis} is defined by
\begin{align}
x \kprod y &\equiv  \nonumber \\
&\left( \frac{x^{\kappa }-x^{-\kappa}}{2}
   + \frac{y^{\kappa }-y^{-\kappa }}{2} +\sqrt{1+\left( \frac{
  x^{\kappa }-x^{-\kappa }}{2}+
  \frac{y^{\kappa }-y^{-\kappa }}{2}\right) ^{2}}
  \right)^{\frac{1}{\kappa }},
\label{def of kappa-product}
\end{align}
which has the following properties 
\begin{eqnarray}
\textrm{associativity} \quad (x \kprod y) \kprod z &=& x \kprod (y \kprod z),
\\
\textrm{unit element} \quad \quad x \kprod 1 &=& 1 \kprod x = x,
\\
\textrm{inverse element} \quad x \kprod (1/x) &=& (1/x) \kprod x = 1,
\end{eqnarray}
and satisfying the following relations
\begin{eqnarray}
\kln \left( x \kprod y\right) &=& \kln (x) +\kln (y),  
\label{property of ln_kappa}
\\
\kexp (x) \kprod \kexp (y)  &=& \kexp (x + y).
\end{eqnarray}
%
%
In this way the $\kappa$-product also has suitable properties for
generalizing Gauss' law of error.
It is then natural to ask whether we can generalize
Gauss' law of error based on the $\kappa$-product.
This paper presents a positive answer to the above question.
We show the $\kappa$-generalization of
Gauss' law of error, in which the $\kappa$-generalized
maximum likelihood principle leads to the $\kappa$-generalized
Gaussian distribution.
In the next section, we present the $\kappa$-generalized version
of the maximum entropy principle in order to derive 
the $\kappa$-Gaussian distributions.
In section 3, by utilizing the $\kappa$-product, 
the likelihood function in Gauss' law of error is generalized. 
We then prove the solution of the $\kappa$-generalized maximum likelihood
principle is the $\kappa$-Gaussian
distribution. The last section is devoted to our concluding remarks.

\section{Maximum $\kappa$-entropy principle}
\vspace*{-7mm}
Kaniadakis' $\kappa$-entropy \cite{Kaniadakis01} can be 
defined \cite{Naudts04} by
\begin{eqnarray}
    \Sk &\equiv& c_{\kappa} \left( 
              1-\int dx f(x)^{1+\kappa} \right) +
              c_{-\kappa} \left(
              1-\int dx f(x)^{1-\kappa} \right), \\
      &\textrm{with}& \quad c_{\kappa} =  
           \frac{1}{2 \kappa (1+\kappa)}.
  \label{kappa-entropy}
\end{eqnarray}
$\Sk$ reduces to Boltzmann-Gibbs-Shannon entropy 
$S^{\rm BGS}=-\int dx f(x) \ln f(x)$
in the limit of $\kappa \to 0$.\\ 
Maximizing $\Sk$ under the two constraints,
\begin{equation}
    \int_{-\infty}^{\infty} dx \; x^2 f(x) = \sigma^2, \quad
    \int_{-\infty}^{\infty} dx f(x) = 1,
\end{equation}
leads to 
\begin{equation}
    \frac{\delta}{\delta f(x)} \left( \Sk - 
        \beta \int_{-\infty}^{\infty} dx \; x^2 f(x) 
     -\gamma \int_{-\infty}^{\infty} dx f(x) \right)= 0,
\end{equation}
where $\beta$ and $\gamma$ are Lagrange multipliers associated
with the two constraints, respectively.
It's solution is the {\it so-called} $\kappa$-Gaussian distribution,
\begin{equation}
    f(x) = \kexp \left( -\beta x^2 - \gamma \right).
    \label{k-Gdf}
\end{equation}
Note that $\gamma$ cannot be factored out from the argument of
the $\kappa$-exponential function unless $\kappa=0$. This is one
of the most different properties of the $\kappa$-exponential
function against the $q$-exponential function.
In fact, applying the similar argument to the $q$-deformed functions,
we can obtain the $q$-Gaussian function 
$f_q(x) = \qexp \left( -\beta' x^2 - \gamma' \right)$, where
two Lagrange multiplier $\beta'$ and $\gamma'$ are introduced. 
Then, due to the property of 
Eq. (\ref{q-identity}), $\gamma'$ can be factored out as
\begin{eqnarray}
    f_q(x) = \qexp \left( -\beta' x^2 - \gamma' \right)
    = \qexp(-\gamma') 
        \qexp\left( \frac{-\beta' x^2}{1-(1-q) \gamma'} \right).
\end{eqnarray}

\begin{figure}[h]
  \begin{center}
    \resizebox{130mm}{!}{\includegraphics{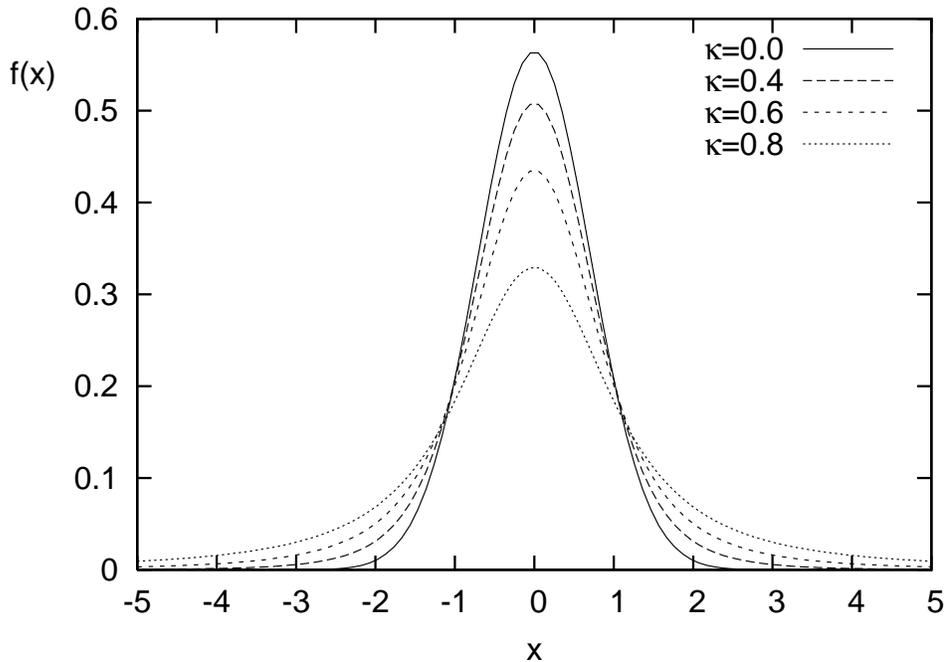}}
    \caption{$\kappa$-Gaussian functions Eq. (\ref{k-Gdf}) for some
     values of $\kappa$ with $\beta=1$. For each curve $\gamma$ is 
determined so that the normalization is satisfied. $\kappa=0$ corresponds 
to the standard Gaussian function.}
    \label{k-Gaussian}
  \end{center}
\end{figure}

In Fig. \ref{k-Gaussian}, $\kappa$-Gaussian functions are plotted
for some different $\kappa$ values.

\section{$\kappa$-generalization of Gauss' law of error}
\vspace*{-7mm}
Let us consider the same situation as conventional Gauss' law of 
error \cite{Jaynes,Suyari}, i.e., 
we get $n$ observed values:
\begin{equation}
x_{1},x_{2},\cdots ,x_{n}\in \mathbb{R}
\end{equation}%
as the results of $n$ measurements for certain observations. 
Each observed value 
$x_{i}\,\left( i=1,\cdots ,n\right) $ is a result of the measurement of 
identically distributed random variable $X_{i}\,\left( i=1,\cdots ,n\right)$. 
There exists a true value $x$ satisfying the \textit{additive}
relation:
\begin{equation}
  x_{i}=x+e_{i}\quad \left( i=1,\cdots ,n\right),  
  \label{q-relation}
\end{equation}%
where each of $e_{i}$ is an error in each observation of the true value $x$.
Thus, for each $X_{i}$, there exists a random variable $E_{i}$ such that $%
X_{i}=x+E_{i}\,\left( i=1,\cdots ,n\right) $. Every $E_{i}$ has the same
probability density function $f$ which is differentiable, because 
$X_{1},\cdots ,X_{n}$ are identically distributed random variables 
(i.e., $E_{1},\cdots ,E_{n}$ are also so).\\
In order to prove the theorem for the $\kappa$-generalization
of Gauss' law of error, we use the following lemma.
Although the proof, which can be found in Ref. \cite{Suyari}, is 
simple and compact, we here show it for the sake of being self-contained. 

\noindent
\textbf{Lemma}
Let $\phi $ be a continuous function from $\mathbb{R}$ into itself and
satisfying that $\sum_{i=1}^{n}\phi \left( e_{i}\right) =0$ for every $n\in
\mathbb{N}$ and $e_{1},\cdots ,e_{n}\in \mathbb{R}$ with $%
\sum\nolimits_{i=1}^{n}e_{i}=0$. Then there exists $a \in \mathbb{R}$ such
that $\phi \left( e\right) = a e.$

\begin{proof}
In the case $n=2$, we can easily see that $\phi \left( -e\right) =-\phi
\left( e\right) $ for every $e = e_1, e_2 \in \mathbb{R}.$ 
Moreover, in the case that $n=3$ 
we have $\phi \left( e_{1}+e_{2}\right) =\phi \left( e_{1}\right) +\phi
\left( e_{2}\right) $ for every $e_{1},e_{2}\in \mathbb{R}$. From this
result and continuity of $\phi(e)$, it is easy to show that $\phi(e)$ must 
be a linear function of $e$, which prove the lemma.
\end{proof}

\noindent
\textbf{Theorem}
For a given set of the data $x_{1},x_{2},\cdots ,x_{n} $,
if the likelihood function 
$L_{\left\{ \kappa \right\} }\left( \theta \right) $ of a
variable $\theta$, which is defined by
\begin{eqnarray}
  L_{\left\{ \kappa \right\} }\left( \theta \right) 
 &=&L_{\left\{ \kappa \right\} }\left( x_{1},x_{2},\cdots ,x_{n};\theta \right)
\nonumber \\ 
 &\equiv& f\left(\theta-x_{1} \right) \kprod f\left( \theta - x_{2}
\right) \kprod \cdots \kprod
f\left( \theta - x_{n} \right),  
\label{kappa-likelihood function}
\end{eqnarray}
takes the maximum at
\begin{equation}
\theta =\theta ^{\ast } \equiv \frac{1}{n} \sum_{i=1}^{n} x_i,
\label{kappa-max value}
\end{equation}
then the probability density function $f$ must be a $\kappa$-Gaussian 
distribution:
\begin{equation}
f\left( \theta-x_i \right) = \exp _{\left\{ \kappa \right\} }
    \Big( -a_{\kappa} (\theta - x_i)^2 +C_{\kappa } \Big).  
\label{kappa}
\end{equation}
where $a_{\kappa}$ is a $\kappa$-dependent positive
constant, and $C_{\kappa }$ is a $\kappa$-dependent 
normalization factor.

\begin{proof}
Taking the $\kappa$-logarithm of the both side of the likelihood function 
$L_{\left\{ \kappa \right\} }\left( \theta \right)$ 
in Eq. \eqref{kappa-likelihood function} leads to
\begin{equation}
 \kln \Big( L_{\left\{ \kappa \right\} }\left( \theta \right) \Big)  
  = \sum_{i=1}^{n} \kln \Big( f( \theta - x_{i}) \Big),
\label{kappa-logLF}
\end{equation}
where the property (\ref{property of ln_kappa}) is used. Differentiating the
above formula Eq. \eqref{kappa-logLF} with respect to $\theta $, we have%
\begin{equation}
\frac{d}{d\theta } \kln 
 \Big( L_{\left\{ \kappa \right\} } \left( \theta \right) \Big) 
= \sum_{i=1}^{n} \frac{d}{d\theta }
  \kln \Big( f(\theta - x_{i}) \Big).  
\label{q-bibun LLF}
\end{equation}
When $\theta =\theta ^{\ast }$ the likelihood function 
$L_{\{ \kappa\}}(\theta)$ takes the maximum, so that
\begin{equation}
  \sum_{i=1}^{n} \frac{d}{d \theta} \ln_{\{ \kappa \}} f(\theta - x_{i}) 
   \Big\vert_{\theta=\theta^{\ast}}=0.
   \label{q-LLF=0}
\end{equation}
Let $e_i$ and $\phi _{\kappa } (e) $ be defined by
\begin{equation}
   e_i \equiv \theta-x_i, \quad \left( i=1,\cdots ,n\right), \quad
  \phi _{\kappa } (e) \equiv
        \frac{d}{d \theta} 
       \ln _{\left\{ \kappa \right\} } f(e) \Big\vert_{\theta=\theta^{\ast}},
\label{q-def e_i}
\end{equation}
respectively.
Then Eq. \eqref{q-LLF=0} can be rewritten to
\begin{equation}
  \sum_{i=1}^{n} \phi _{\kappa }(e_i) =0.
\label{q-condition1}
\end{equation}
On the other hand, from Eq. (\ref{kappa-max value}) , we obtain%
\begin{equation}
  \sum_{i=1}^{n} \left( \theta^{\ast} -x_{i} \right)
  = \sum_{i=1}^{n} e_i =0.
\label{q-condition2}
\end{equation}
Now our problem is reduced to determining the function $\phi _{\kappa }(e)$
simultaneously satisfying Eqs. (\ref{q-condition1}) and (\ref{q-condition2}).
By using the lemma, $\phi _{\kappa }(e)$ should be proportional
to $e$, i.e.,
\begin{equation}
  \phi _{\kappa }(e) = -2 a_{\kappa} e,
\label{prop-rel} 
\end{equation}
where $a_{\kappa}$ is a $\kappa$-dependent positive constant,
and the factor $-2$ is introduced for the sake of the simplicity of the
final result.
Integrating Eq. (\ref{prop-rel}) w.r.t $e$ gives
\begin{equation}
  \kln f(e) = -a_{\kappa} e^2 + C_{\kappa},
\end{equation}
where $C_{\kappa}$ is a normalization factor which
is $\kappa$-dependent through $a_{\kappa}$. 
Thus we obtain
\begin{equation}
  f(e) = \kexp \Big( -a_{\kappa} e^2 + C_{\kappa} \Big).
\end{equation}
A normalizable distribution requires that $a_{\kappa}$ should be positive,
and the normalization then determines the factor $C_{\kappa}$.

Finally let us confirm that the extremum of the likelihood function 
at $\theta=\theta^{\ast}$ is the maximum.
By using Eq. (\ref{prop-rel}), we obtain
\begin{eqnarray}
\frac{d^2}{d\theta^2} \kln \Big( 
   L_{\{ \kappa \}}(\theta)\Big) &=&
   \frac{d}{d \theta} \left(
 \frac{d}{d \theta}  \phi_{\kappa}(\theta -x_i) \right)
\nonumber \\
  &=& \frac{d}{d \theta} \left( -2 a_{\kappa} \dot(\theta-x_i) \right)
= -2 a_{\kappa}.
\end{eqnarray}
Since $a_{\kappa}$ is positive, 
$L_{\left\{ \kappa \right\} }\left( \theta^{\ast} \right)$ is the maximum. 
\end{proof}

\section{Concluding remarks}
\vspace*{-7mm}
We have shown two different methods to derive the $\kappa$-Gaussian
distribution, i.e., both maximum entropy and maximum likelihood principles 
are generalized by utilizing $\kappa$-deformed
functions and associated multiplication operation ($\kappa$-product). 
Both $\kappa$-generalized methods reduce to the standard ones, respectively, 
in the limit of $\kappa \to 0$. As the $q$-product is the most fundamental
ingredient in the $q$-generalization of Gauss' law of error \cite{Suyari},
the $\kappa$-product enables us generalizing Gauss' law of error for 
the $\kappa$-Gaussian.

As seen in both $q$- and $\kappa$-generalizations of 
the maximum likelihood principle, we remark that for any set 
of one-parameter generalizations of exponential
and logarithmic functions we can generalize the likelihood function by
introducing a suitable multiplication operation or product. 
%
Let us here denote this product as $\aprod$, where $a$ stands
for a real-parameter of the deformed functions.
By using this product we can define
the generalized likelihood function, which is similar to 
Eq. \eqref{kappa-likelihood function} but $\kappa$-products are replaced
with the products $\aprod$. The associated Gaussian distribution can be
obtained from this $a$-parameter generalization of Gauss' law of error. 

The difference among $q$-, $\kappa$-, and other generalizations of
Gauss' law of error apparently lies in the definition of each generalized
likelihood function in terms of the generalized products. 
For the standard case, 
the standard products of each distribution $f(\theta-x_i)$ define
the original Gauss likelifood function $L(\theta)$, which is the
$\kappa \to 0$ limit of Eq. \eqref{kappa-likelihood function}.
It is well-known that the standard products of $f$ means 
all $f$ are statistically independent.
Unfortunately until now there is no consensus
for interpretating any generalized likelihood function nor
generalized product, although we feel that a certain kind of correlation
is relevant \cite{Tsallis05}.

\end{document}